\documentclass[11pt,twoside]{article}
\usepackage{atmp}
\copyrightnotice{2001}{5}{947}{968}    %% year, volume, first page, last page.

\usepackage{graphicx}

\newcommand{\va}{\scriptscriptstyle}

\begin{document}

\title{Spin foam quantization of SO(4) Plebanski's action}

\url{abs/gr-qc/0203058}        %% Everything after http://xxx.lanl.gov/

\author{A. Perez}       %% Can have multiple \author, \address
 \address{Center for Gravitational Physics and Geometry\\
 Pennsylvania State University\\ University Park, PA}
 \addressemail{perez@gravity.psu.edu}  %% Under \address only, please!

 \markboth{\it Spin foam quantization \ldots}{\it A. Perez}

\begin{abstract}
% Body of abstract.
The goal of this work is two-fold. In the first part of this paper
we regard classical Plebanski's action as a BF action supplemented
by constraints (defined in the spirit of Barrett and Crane). We
introduce a spin foam model for Riemannian general relativity by
systematically implementing these constraints as restrictions on
paths in the state-sum of the BF theory. The spin foam model
obtained is closely related to ---but not the same as--- the
Barrett-Crane model. More precisely, configurations satisfying our
constraints correspond to a subset of the Barrett-Crane
configurations. Surprisingly, all tetrahedra in the allowed
configurations turn out to have zero volume.

In the second part of the paper we study the quantization of the
effective action corresponding to the degenerate sectors of
Plebanski's theory and obtain a very simple spin foam model. This
model turns out to be precisely the one introduced by De Pietri et
al. as an alternative to the one proposed by Barrett and Crane.
This result establishes a clear-cut connection of the model with a
classical action. The 4-simplex configurations of the model
corresponding to the full Plebanski's action (obtained in the
first part) turn out to be entirely contained in the set of
configurations of the model of the degenerate sector.
 \end{abstract}

 \cutpage
 \section{Introduction}
% First section text.
% . . .
% End of paragraph that first fills page.
%
In reference \cite{BC2} Barrett and Crane introduced a very
interesting model of Riemannian quantum gravity based on a
constrained state-sum. The definition of the model can be 
nicely motivated by geometrical
properties of the so-called `quantum tetrahedron'. The definition
of the quantum tetrahedron in 3 dimensions was originally
introduced by Barbieri in \cite{barb2}. Baez and Barrett showed
that a generalization to 4 dimensions naturally leads to the the
Barrett-Crane (BC) model \cite{baez7, baez6}. Evidence suggesting
that the model corresponds to a discrete path integral for general
relativity has been found in \cite{bawi,crane}. The model turns
out to be well defined on a finite (non-degenerate) triangulation
once an appropriate normalization is chosen\cite{a7}. This
normalization arises naturally in the so-called group field theory
(GFT) formulation \cite{reis1,reis2}, which in addition provides a
prescription for summing over discretizations. The model has been
extended to the Lorentzian sector in \cite{BC1,a9,a8}. The
finiteness properties are preserved in this extension\cite{a2}.

The $SO(4)$ Plebanski action corresponds to the $SO(4)$ BF action
plus certain Lagrange multiplier terms imposing constraints on the
$B$ field. Therefore, one can formally quantize the theory
restricting the BF-path-integral to paths that satisfy the B-field
constraints. In the literature, there is an implicit assumption
that the BC model corresponds to a realization of this idea. In
other words, the definition of the quantum tetrahedron in 4d
(giving rise to the BC model) is sometimes regarded as an
alternative way to impose the required restrictions on the
B-configurations in the discretization. The purpose of this work
is to analyze if this is the case by systematically
carrying out this restrictions.

We will present a construction which defines the path integral of
Plebanski's action on a fixed simplicial decomposition of
space-time. As just mentioned, this is done by appropriately
restricting the state-sum of the $SO(4)$ BF theory. The
path-integral of the BF theory is defined on a triangulation using
techniques similar to those in lattice-gauge theory. The spin-foam
formulation ---or state-sum--- is obtained by performing the mode
expansion of certain distributions on $SO(4)$. This is analogous
to a Fourier transform where modes correspond to unitary
irreducible representations of $SO(4)$ (Peter-Weyl theorem). The
constraints on the B-field in the classical action can be
naturally translated into restrictions on these modes. The
definition of these constraints is not different in spirit from
that of Barrett and Crane. However, we emphasize the requirement
that the restrictions should be imposed on configurations of the
BF theory. After making some natural definitions, a systematic
derivation leads to a model that is closely related to the BC
model but that does not agree with it. This new version has a
puzzling feature: states of 3-geometries (boundary spin-network
states) are annihilated by the volume operator. The point of view
is related to that of Reisenberger and Freidel-Krasnov in
\cite{reis4,reis6,fre5}. No obvious modification of the
prescription can lead to all the BC configurations.

In order to find a possible interpretation of this result we
concentrate on one of the degenerate sectors of Plebanski's action
described in \cite{reis0}. It turns out that one can define a spin
foam quantum model corresponding to this sector in a
straightforward way. For this, one simply applies the same
techniques used in the case of the BF theory. Surprisingly, the
model obtained coincides with the one introduced by De Pietri et
al. in \cite{fre2}. This model was defined as an alternative to
the BC model arising naturally in the context of the group field
theory (GFT) framework. Our result provides a clear-cut
interpretations of the De Pietri et al. formulation as a
quantization of a classical action. An interesting result is that
all the allowed configurations for a 4-simplex in the previous
model (corresponding to generic theory) are special configurations
of this model.

The article is organized in the following way. In the next section
we recall essential facts about $SO(4)$ Plebanski formulation. In
Section \ref{BF} we briefly review the spin foam quantization of
the BF theory and introduce our basic definitions. In Section
\ref{ccc} we solve the constraints that lead one from the BF
theory to general relativity and construct the corresponding
state-sum model. We interpret the results and show that
configurations have zero 3-volume. In Section \ref{ddeegg} we
quantize the effective action corresponding to the degenerate
sectors of Plebanski's action and show that the previous model
corresponds to a sub-set of the spin foam configurations obtained
in the degenerate sector. We end with concluding remarks in
Section \ref{Diss}.

\section{Classical $SO(4)$ Plebanski action}

Let us start by briefly reviewing Plebanski's
formulation\cite{pleb} at the classical level. Plebanski's
Riemannian action depends on an $so(4)$ connection $A$, a
Lie-algebra-valued 2-form $B$ and Lagrange multiplier fields
$\lambda$ and $\mu$. Writing explicitly the Lie-algebra indices,
the action is given by
\begin{equation}\label{pleb}
S[B,A,\lambda,\mu]=\int \left[B^{IJ}\wedge F_{IJ}(A) + \lambda_{IJKL}
\ B^{IJ} \wedge B^{KL} +\mu \epsilon^{IJKL}\lambda_{IJKL} \right],
\end{equation}
where $\mu$ is a 4-form and
$\lambda_{IJKL}=-\lambda_{JLKI}=-\lambda_{IJLK}=\lambda_{KLIJ}$ is
tensor in the internal space. Variation with respect to $\mu$
imposes the constraint $\epsilon^{IJKL}\lambda_{IJKL}=0$ on
$\lambda_{IJKL}$. $\lambda_{IJKL}$ has then $20$ independent
components. Variation with respect to the Lagrange multiplier
$\lambda$ imposes $20$ algebraic equations on the $36$ $B$.
Solving for $\mu$ they are
\begin{equation}
B^{IJ}\wedge B^{KL}-\frac{1}{4!}\epsilon_{OPQR}B^{OP}\wedge
B^{QR}\epsilon_{IJKL}=0
\end{equation}
which is equivalent to
\begin{equation}\label{constraints}
\epsilon_{IJKL} B^{IJ}_{\mu\nu}B^{KL}_{\rho\sigma}=e
 \epsilon_{\mu\nu\rho\sigma},
\end{equation}
for $e\not=0$ where
$e=\frac{1}{4!}\epsilon_{OPQR}B^{OP}_{\mu\nu}B^{QR}_{\rho\sigma}\epsilon^{\mu\nu\rho\sigma}$\cite{fre6}.
The  solutions to these equations are
\begin{equation}\label{ambi}
B=\pm {}^*( e \wedge e), \ \ \ {\rm and}\ \ \ B=\pm e\wedge e,
\end{equation}
in terms of the $16$ remaining degrees of freedom of the tetrad
field $e^I_a$. If one substitutes the first solution into the
original action one obtains an effective action that is precisely
that of general relativity in the Palatini formulation
\begin{equation}\label{pala}
S[e,A]=\int {\rm Tr}\left[e\wedge e \wedge {}^*F(A)\right].
\end{equation}

\section{Quantum $SO(4)$ BF theory}\label{BF}

Classical ($Spin(4)$) BF theory is defined by the action
\begin{equation}
S[B,A]=\int {\rm Tr}\left[B\wedge F(A) \right],
\end{equation}
where $B^{IJ}_{ab}$ is a $Spin(4)$ Lie-algebra valued 2-form,
$A^{IJ}_a$ is a connection on a $Spin(4)$ principal bundle over
$\cal M$. The theory is rather trivial and all classical solutions
are locally equivalent (up to gauge transformations). The theory
has only global degrees of freedom.

One can quantize the theory \`a la Feynman introducing a path
integral measure. This is easily done by replacing the manifold
$\cal M$ by an arbitrary simplicial decomposition $\Delta$
\footnote{More generally, the path integral for the BF theory can
be defined on an arbitrary cellular decomposition of $\cal M$. See
\cite{a1}.}. Take a fixed triangulation $\Delta$ of $\cal M$. The
2-skeleton of the dual of the triangulation defines a cellular
2-complex $\Delta^*$. Associate $B_f\in so(4)$ to each triangle in
$\Delta$ (for convenience we use the face sub index $f$ since
triangles are in one-to-one correspondence to faces $f \in
\Delta^*$), and a group element $g_e \in Spin(4)$ to each edge
$e\in \Delta^*$. Consider the holonomy around faces
$U_f=g_{e_1}g_{e_2}\dots g_{e_n}$, i.e., the product of group
elements of the corresponding edges around one face (an arbitrary
orientation of faces has been chosen). The discretized version of
the partition function becomes
\begin{equation}\label{Zdbf}
{\cal Z}(\Delta)=\int \prod_{f \in \Delta^*} dB^{\va (6)}_f \
\prod_{e \in \Delta^*} dg_e  \ e^{i {\rm Tr}\left[B_f U_f\right]}.
\end{equation}
The measure $dB^{\va (6)}_f$ is the Lebesgue measure on $\Re^6$,
while $dg$ corresponds to the normalized Haar measure of
$Spin(4)$. Now the integration over the $B_f$'s can be done
explicitly \cite{thesis}, and the result is:
\begin{equation}\label{papart}Z(\Delta)= \int \prod_{e \in \Delta^*}
dg_e \prod_{f \in \Delta^*} \delta(g_{e_1} \cdots g_{e_n}).
\end{equation}

Expanding the delta distribution in unitary irreducible
representations (Peter-Weyl decomposition \footnote{\label{PW}
Peter-Weyl theorem implies that
\[\delta(g)=\sum \limits_{\rho} \Delta_{\rho} {\rm Tr}[\rho(g)]
,\] where $\rho(g)$ is the unitary irreducible representation of
dimension $\Delta_{\rho}$.}) we obtain
\begin{equation}\label{coloring}
{\cal Z}(\Delta)=\sum \limits_{ {\cal C}:\{\rho\} \rightarrow
\{f\} } \int \ \prod_{e \in \Delta^*} dg_e  \
\prod_{f \in \Delta^*} \Delta_{\rho_f} \ {\rm
Tr}\left[\rho_f(g^1_e\dots g^{\va N}_e)\right],
\end{equation}
where ${\cal C}:\{\rho\} \rightarrow \{ f \}$ denotes the
assignment of irreducible representations to faces in the dual
2-complex $\Delta^*$. Each particular assignment is
referred to as a {\em coloring}, $\cal C$.

Next step is to integrate over the connection $g_e$. Since edges
$e\in\Delta^*$ bound four different faces, each group element $g_e$
appears in the mode expansion of four delta functions in
(\ref{papart}). The formula we need is that of the projection
operator into the trivial component of the tensor product of
four irreducible representations, namely
\begin{equation}\label{4dp}
\int dg\ {\rho_1(g)}\otimes \rho_2(g) \otimes \rho_3(g) \otimes
\rho_3(g)= \sum_{\iota} {C^{\va \iota}_{\va \rho_1 \rho_2 \rho_3
\rho_4} \ C^{*{\va \iota}}_{\va \rho_1 \rho_2 \rho_3 \rho_4}},
\end{equation}
where $C^{\va \iota}\in {\cal H}_{\rho_1}\otimes {\cal H}_{\rho_2}
\otimes {\cal H}_{\rho_3} \otimes {\cal H}_{\rho_4}$ represents an
orthonormal basis of invariant vectors and the sum on the RHS
ranges over all the basis elements $\iota$.

The RHS of equation (\ref{4dp}) can be
represented graphically as
\begin{eqnarray}\label{4dp}
\sum_{\iota} {C^{\va \iota}_{\va \rho_1 \rho_2 \rho_3 \rho_4} \
C^{*{\va \iota}}_{\va \rho_1 \rho_2 \rho_3 \rho_4}}= \sum_{\iota}
\ \begin{array}{c}
\includegraphics[width=2.3cm]{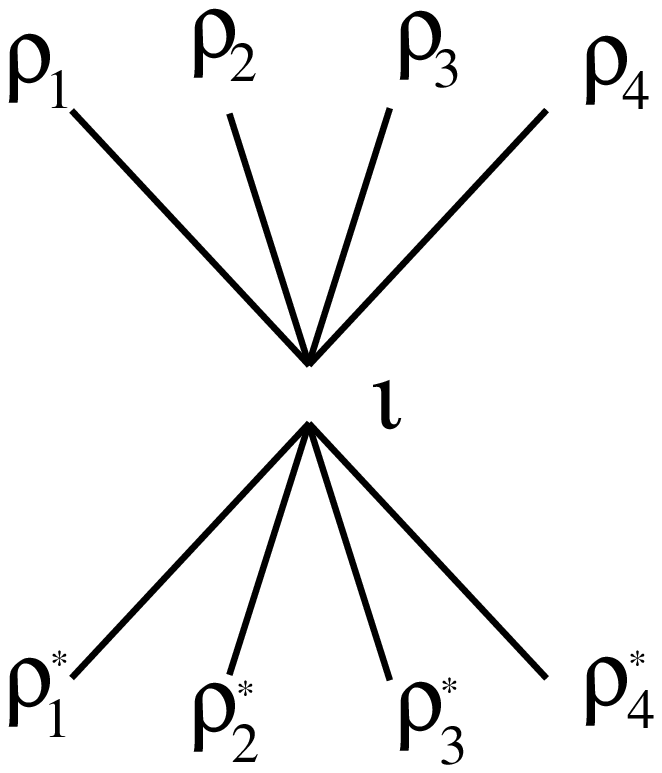}\end{array}.
\end{eqnarray}
Therefore, integrating over the $g_e$'s using (\ref{4dp}) and
keeping track of indices we obtain
\begin{eqnarray}\label{bf4} Z_{BF}(\Delta)=\sum \limits_{ {\cal
C}_f:\{f\} \rightarrow \rho_f }  \sum \limits_{{\cal C}_e:\{e\}
\rightarrow \{ \iota_e \}} \ \prod_{f \in \Delta^*} \Delta_{\rho}
\prod_{v \in {\Delta^*}}
\begin{array}{c}
\includegraphics[width=3cm]{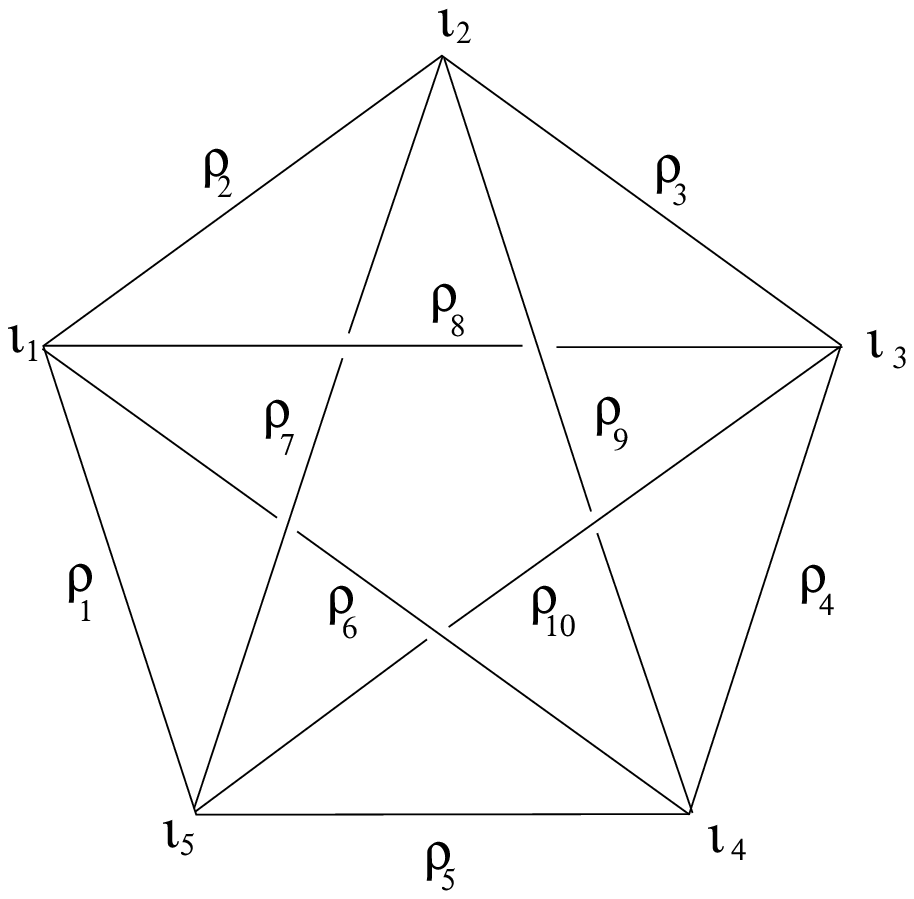}\end{array},
\end{eqnarray}
where the pentagonal diagram representing the vertex amplitude
denotes the trace of the product of five intertwiners $C^{\va
\iota}_{\va \rho_1 \rho_2 \rho_3 \rho_4}$ according to the
graphical notation of (\ref{4dp}). Vertices $v\in \Delta^*$ are in
one-to-one correspondence to 4-simplexes in the triangulation
$\Delta$. In addition we also have ${\cal C}_e:\{e\} \rightarrow
\{ \iota_e \}$ representing the assignment of intertwiners to
edges. The sum over the {\em coloring} of edges, ${\cal C}_e$,
comes from (\ref{4dp}) (for an extensive explanation of the
construction of the state-sum for the BF theory and the notation
used here see\cite{baez5}).

What happened in going from equation (\ref{Zdbf}) to
(\ref{coloring})? We have replaced the continuous multiple
integral over the $B$'s by the sum over representations of
$SO(4)$. Roughly speaking, the degrees of freedom of $B$ are now
encoded in the representation being summed over in
(\ref{coloring}). One can make a more precise definition of what
`$B$' is at the level of (\ref{coloring}). In order to motivate
our definition we isolate a single face contribution to the
integrand in the partition function (\ref{Zdbf}). Then we notice
that the right invariant vector field $-i{\cal X}^{IJ}(U)$ has a
well defined action at the level of equation (\ref{coloring}) and
acts as a `quantum' B at the level of (\ref{Zdbf}) since
\begin{eqnarray}\label{RIV}
\nonumber& &-i{\cal X}^{IJ}(U)\left( e^{i{\rm
Tr}[BU]}\right)|_{U\sim 1} = X^{IJ\ \mu }_{\ \ \ \ \ \nu}U^{\nu}_{\ \
\sigma}\frac{\partial}{\partial
U^{\mu}_{\ \ \sigma}} e^{i{\rm Tr}[BU]}|_{U\sim 1}= \\
&& ={\rm Tr}[X^{IJ}UB] e^{i{\rm Tr}[BU]}|_{U\sim 1}\sim B^{IJ}e^{i{\rm Tr}[BU]},
\end{eqnarray}
where $X^{IJ}$ are elements of an orthonormal basis in the $SO(4)$
Lie-algebra. The evaluation at $U=1$ is motivated by the fact that
configurations in the BF partition function (\ref{papart}) have
support on flat connections.

The constraints in (\ref{constraints}) are quadratic in the $B$'s.
We have then to worry about
cross terms, more precisely the nontrivial case corresponds to:
\begin{eqnarray}\label{RIV2}&& \nonumber
\epsilon_{IJKL}{\cal X}^{IJ}(U){\cal X}^{KL}(U)\left( e^{i{\rm
Tr}[BU]}\right)|_{U\sim 1}\\ \nonumber && =
-\epsilon_{IJKL} \left({\rm Tr}[X^{IJ}UB]{\rm Tr}[X^{KL}UB] e^{i{\rm Tr}[BU]}
+i{\rm Tr}[X^{IJ}X^{KL}UB]e^{i{\rm Tr}[BU]}\right)|_{U\sim 1}\\ &&
\sim \epsilon_{IJKL}B^{IJ}B^{KL}e^{i{\rm Tr}[BU]},
\end{eqnarray}
where the second term on the second line can be dropped using that
$\epsilon_{IJKL} X^{IJ}$\newline\noindent$X^{KL}\propto 1$ (one of the two $SO(4)$
Casimir operators) and $U\sim 1$. Therefore, we define the $B_f$
field associated to a face at the level of equation
(\ref{coloring}) as the appropriate right invariant vector field
$-i{\cal X}^{IJ}(U_f)$ acting on the corresponding discrete
holonomy $U_f$, namely
\begin{equation}\label{B_f}
B_f^{IJ}\rightarrow -i{\cal X}^{IJ}(U_f).
\end{equation}
It is easy to verify that one can use left invariant
vector fields instead in the previous definition without
changing the following results.

\section{Implementation of the constraints that reduce the BF theory to general relativity}\label{ccc}

\subsection{Formulation of the problem}

\begin{figure}
\centering{\includegraphics[width=5cm]{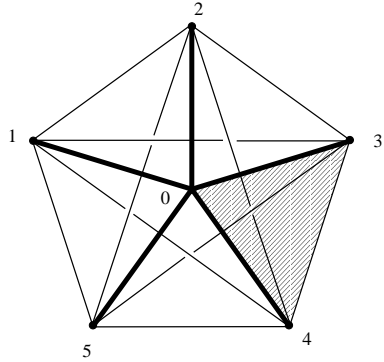}}
\caption{A fundamental {\em atom} is defined by the
intersection of a dual vertex in $\Delta^*$ (corresponding to a 4-simplex in $\Delta$)
with a 3-sphere. The thick lines represent the internal edges while the thin lines the
intersections of the internal faces with the boundary. One of the
faces has been emphasized.}\label{chunk}
\end{figure}

Now we describe the implementation of the constraints
(\ref{constraints}). The idea is to concentrate on a single
4-simplex amplitude using the locality of the BF theory state sum
\footnote{The term `local' here is used as defined by Reisenberger
in \cite{reis4}. It means that the spin foam can be written as
4-simplex contributions that communicate with other 4-simplexes by
boundary data (connection). The full amplitude is obtained by
integrating out the boundary connections along the common boundary
of the 4-simplexes that make up the simplicial complex.}. The
4-simplex wave function is obtained using (\ref{papart}) on the
dual 2-complex with boundary defined by the intersection of the
dual of a single 4-simplex with a 3-sphere, see Figure
\ref{chunk}. We refer to this fundamental building block as `atom'
as in \cite{reis4}.
The boundary values of the discrete connection are held fixed. We
denote as $h_{ij}\in Spin(4)$ ($i\not=j$, $i,j=1\cdots 5$ and $h_{ij}=h^{-1}_{ji}$) the
corresponding 10 boundary variables (associated to thin boundary
edges in Figure \ref{chunk})
\footnote{Strictly speaking, the boundary connections $h_{ij}$ are
defined as the product $h^{\prime}_{ij} h^{''}_{ij}$ where
$h^{\prime}$ and $h^{''}$ are associated to half paths as follows:
take the edge $ij$ for simplicity and assume it is oriented from
$i$ to $j$. Then $h^{\prime}_{ij}$ is the discrete holonomy from
$i$ to some point in the center of the path and $h^{''}_{ij}$ is
the holonomy from that center point to $j$. This splitting of
variables is necessary when matching different atoms to
reconstruct the simplicial amplitude. The use of this variables
(wedge variables) will be crucial in Section \ref{ddeegg}. For a
more detailed description of wedge variables see
\cite{reis4,reis6}.}
and $g_i\in Spin(4)$ ($i=1,\cdots, 5$) the internal connection (corresponding
to the thick edges in Figure \ref{chunk}). According to (\ref{papart}) the 4-simplex BF amplitude
$4SIM_{BF}(h_{ij})$ is given by
\begin{equation}\label{pito}
4SIM_{BF}(h_{ij})=\int \prod_i dg_i  \prod_{i<j} \delta(g_ih_{ij}g_j).
\end{equation}
With the definition of the $B$ fields given in (\ref{B_f}) the
constrained amplitude, $4SIM_{const}(h_{ij})$, formally becomes
\begin{equation}
4SIM_{const}(h_{ij})=\int \prod_i dg_i \delta\left[{\rm Constraints}({\cal X}(U_{ij}))\right]\prod_{i<j} \delta(g_ih_{ij}g_j),
\end{equation}
where $U_{ij}=g_ih_{ij}g_j$ is the holonomy around the triangular face (wedge) $0ij$
according to Figure \ref{chunk}.
It is easy to verify, using an equation analogous to (\ref{RIV}), that one can
define the $B$'s by simply acting with the right invariant vector fields on the
boundary connection $h_{ij}$. Therefore, the previous equation is equivalent to
\begin{equation}\label{4sgr}
4SIM_{const}(h_{ij})= \delta\left[{\rm Constraints}({\cal X}(h_{ij}))\right] \int \prod_i dg_i \prod_{i<j} \delta(g_ih_{ij}g_j),
\end{equation}
where we have taken the delta function out of the integral. The
quantity on which the formal delta distribution acts is simply
$4SIM_{BF}(h_{ij})$ (defined in (\ref{pito})), which after
integrating over the internal connection $g_i$, and using equation
(\ref{4dp}) becomes
\begin{equation}\label{bfwf}
\!\!4SIM_{BF}(h_{ij})=\!\!\!\!\sum \limits_{\rho_1 \cdots \rho_{10}}\! \sum
\limits_{\iota_1 \cdots \iota_5}\!\!
\begin{array}{c}
\includegraphics[width=3cm]{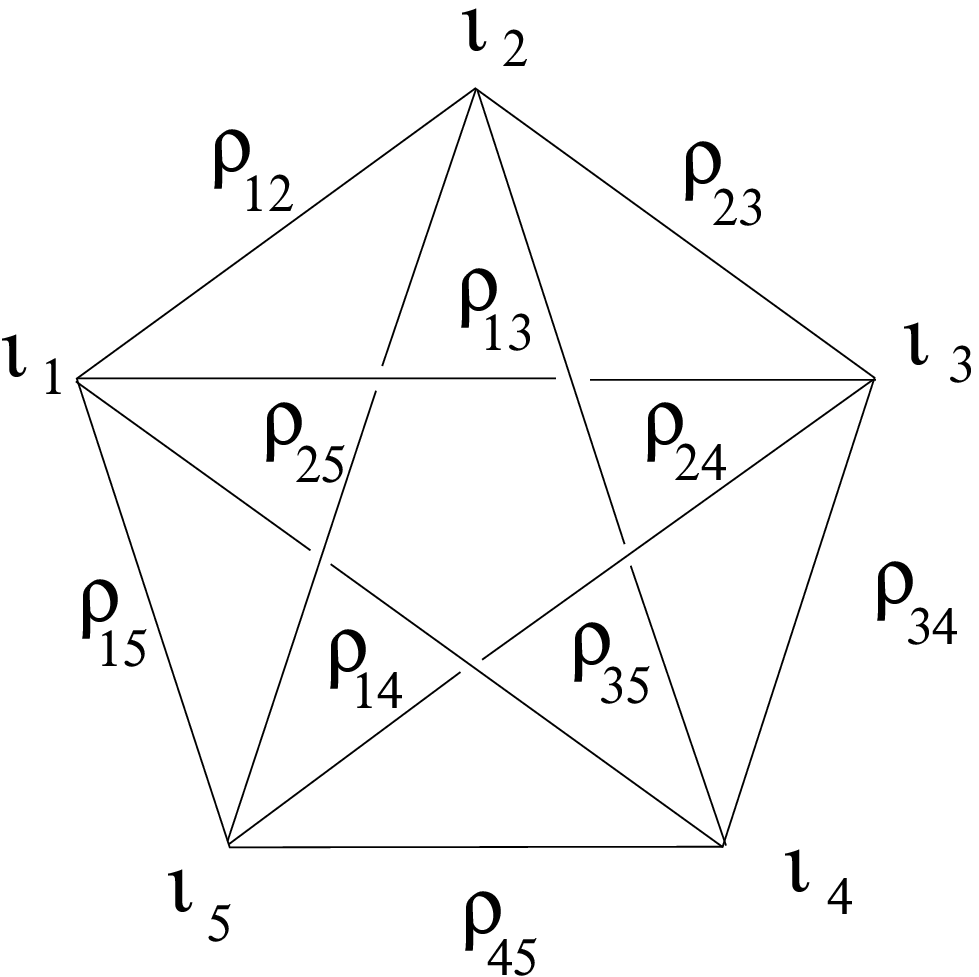}\end{array}
\begin{array}{c}
\includegraphics[width=4.5cm]{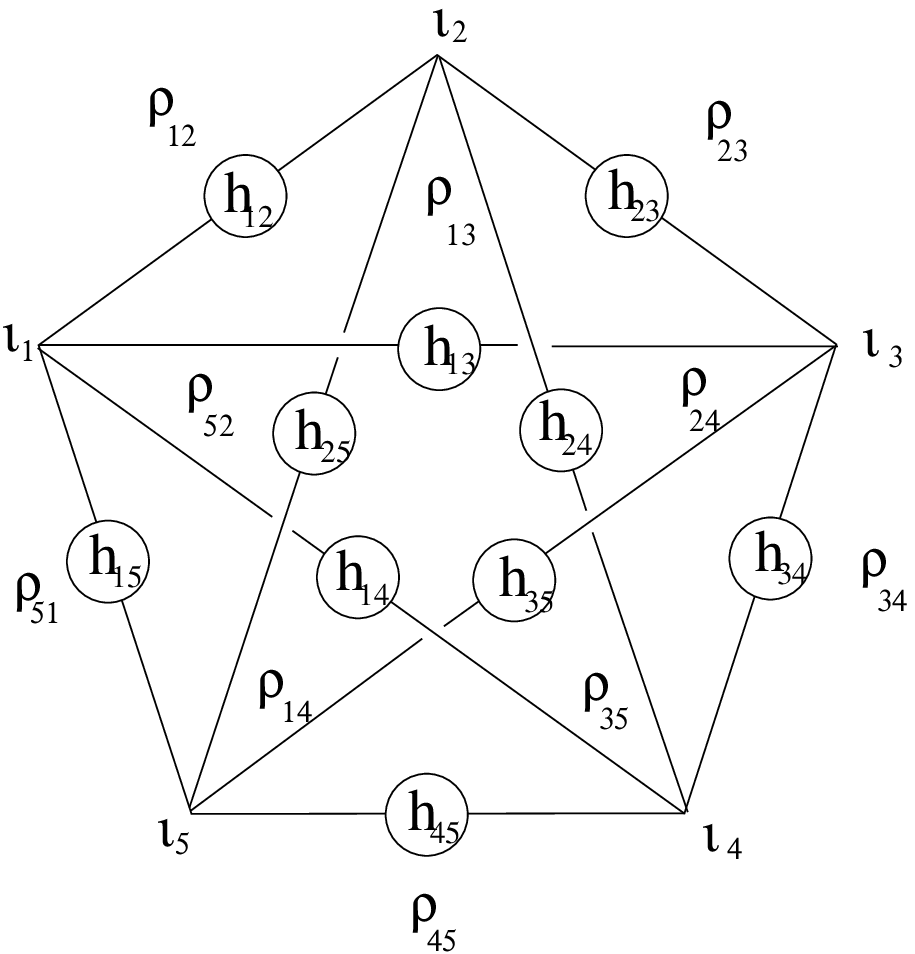}\end{array},
\end{equation}
where the circles represent the corresponding $\rho$-representation
matrices evaluated on the the corresponding boundary connection $h$.
The term on the left is a $15j$-symbol as in (\ref{bf4}) while the term
on the right is the trace of five intertwiners with the respective
boundary connection insertions. Notice then that nodes on the two
pentagonal diagrams are linked together by the value of their intertwiner.

The 4-simplex amplitude for the constraint spin foam model is then defined as the
restriction of $4SIM_{BF}(h_{ij})$ imposed by the quantum version
of the constraints (\ref{constraints}). The latter are defined by the
following set of differential equations
\begin{equation}\label{XX}
\epsilon_{IJKL}{\cal X}^{IJ}(h_{ij}){\cal X}^{KL}(h_{ik})\
4SIM_{const}(h_{ij})= 0 \ \ \ \forall \ j,k,
\end{equation}
and where the index $i=1,\cdots, 5$ is held fixed. The translation
of the continuum constraint (\ref{constraints}) into discrete
elements associated to faces in $\Delta^{*}$ is analogous to that
given in \cite{baez7,fre6}. Notice that (\ref{XX}) is to be
thought as a condition on BF amplitudes and is not a general
equation to be imposed to any 4-simplex amplitude. Recall that the
strategy is to constraint the BF theory to obtain a definition of
the path integral for general relativity
\footnote{\label{fn}We illustrate the general idea with
the following simple example. Imagine that the analog of $4SIM_{BF}$
function (eq. (\ref{pito})) is the integral
\begin{equation}\label{minisf}
A=\int dk dp  e^{ik x+ip y}=\delta(x)\delta(y),
\end{equation}
where $x, y \in [0,2\pi]$ represent the boundary `connections'. The analog of the constraint
(\ref{constraints}) is defined to be $k-p=0$ which in turn implies the constrained
amplitude to be
\[A_{const}=\delta(x+y).\]
Let us now apply the prescription used in the BF theory. We can
expand the un-constrained function (\ref{minisf}) in terms of
`spin foam' amplitudes
\[A=\frac{1}{4\pi^2}\sum \limits_{n,m} e^{i n x+i m y}.\]
In this case this corresponds to Fourier expanding delta function
on $S_1 \times S_1 $ (Peter-Weyl decomposition for $U(1) \times
U(1)$). The constraint is now represented by a combination $C$ of
right invariant vector fields on $U(1)$:
$C=\partial_x-\partial_y$. So we can now impose the constraints by
means of selecting those configurations (modes) in (\ref{minisf})
that are annihilated by $C$. The equation analogous to (\ref{XX})
is
\[ (\partial_x-\partial_y)e^{i n x+i m y}=(n-m)e^{i n x+i m y}=0\]
which implies $n=m$ and $A_{const}=\delta(x+y)$. }.

Equations closely related to (\ref{XX}) can also be obtained as
the geometric restrictions on the $B$'s to be simple bi-vectors
coming from a dual cotetrad or to characterize the geometry of a
tetrahedron in 4 dimensions \cite{baez6,baez7}. In this case the
equivalent of $B$ correspond to bivectors defined by the faces of
a classical tetrahedron. Using geometric quantization one obtains
the Hilbert space of states of the `quantum tetrahedron' where the
$B$'s are promoted to operators. Notice the our $B$ operator
(\ref{RIV}) is obtained directly from the BF path integral and one
does not need to invoke any additional quantization principle. The
procedure is completely analogous to the simple example of
Footnote \ref{fn}. A similar point of view has been taken by
Reisenberger and Freidel-Krasnov in \cite{reis4, fre5}.

\subsection{Restricted BF paths}

The following procedure is very similar in spirit to the BC
prescription\cite{baez7,baez6}. The essential difference is that
we now require the set of restricted configurations to be
contained in the set of modes of the BF amplitude, $4SIM_{BF}$.

There are seven equations (\ref{XX}) for each value of
$i=1,\cdots,5$. If we consider all the equations for the 4-simplex
amplitude then some of them are redundant. The total number of independent
conditions is 20, in agreement with the number of classical
constraints (\ref{constraints}). For a given $i$ in (\ref{XX})
(i.e., a given tetrahedron) and for $j=k$ the equation becomes
\begin{eqnarray}\label{XX1} \nonumber &&
\epsilon_{IJKL}{\cal X}^{IJ}(h_{ij}){\cal X}^{KL}(h_{ij})\
4SIM_{const}(h_{ij})\\ &&\ \ \ \ \ \ \ \ \ \ \ \ \ \ \ \ \ \
=\left[j_{ij}^{\ell}(j_{ij}^{\ell}+1)-j_{ij}^r(j_{ij}^r+1)\right]
4SIM_{const}(h_{ij})=0,
\end{eqnarray}
where we have used $\rho=j^{\ell}\otimes j^{r}$ for $j^{\ell},
j^{r} \in {\rm Irrep}[SU(2)]$. The previous constraints are solved
by requiring the corresponding representation $\rho_{ij}$ to be
simple, i.e., $\rho_{ij}=j_{ij} \otimes j_{ij}$.

\begin{figure}
\centering{\includegraphics[width=5cm]{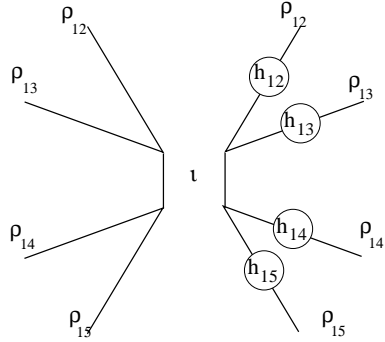}}
\caption{A tree decomposition of one of the nodes in (\ref{bfwf}).
Any tree decomposition is equivalent.}\label{nodo}
\end{figure}

This solves 10 of the 20 equations. The next non-trivial condition
imposed by (\ref{XX}) is when $j\not=k$. In this case we have
\begin{eqnarray}\label{XX2} \nonumber &&
2 \epsilon_{IJKL}{\cal X}^{IJ}(h_{ij}){\cal X}^{KL}(h_{ik})\
4SIM_{const}(h_{ij})\\ && \nonumber =\epsilon_{IJKL} \left({\cal
X}^{IJ}(h_{ij})+{\cal X}^{IJ}(h_{ik})\right) \left({\cal
X}^{KL}(h_{ij})+{\cal X}^{KL}(h_{ik})\right)4SIM_{const}(h_{ij})\\
&&\nonumber
=\left[\iota^{\ell}(\iota^{\ell}+1)-\iota^r(\iota^r+1)\right]
4SIM_{const}(h_{ij})\\ && =0,
\end{eqnarray}
where we used the gauge invariance at the 3-valent node in the
tree decomposition that pairs the representation $\rho_{ij}$ with
the $\rho_{ik}$\footnote{ The gauge invariance at the node allows
us to express the sum of right-invariant vector fields acting on
the external `legs' (see Figure \ref{nodo}.) as a right-invariant
vector field acting on the internal representation $\iota$. Of
course right versus left invariant vector field is a matter of
convention which implies a choice of orientation. Since $Spin(4)$
representations are self dual we have ${\cal X}^{IJ}_{R}(h)=-{\cal
X}^{IJ}_{L}(h^{-1})$.}, and that we have already solved
(\ref{XX1}). In the last line we assume that the internal color of
the corresponding 4-intertwiner is
$\iota=\iota^{\ell}\otimes\iota^{r}$. This choice of tree
decomposition in the case $ij=12$ and $ik=13$ is illustrated in
Figure \ref{nodo}. The solution is clearly
$\iota=\iota\otimes\iota$.

What happens now with any
of the other two remaining conditions, for example,
$E(ij^{\prime}, ik^{\prime})$ for $k\not=k^{\prime}$, $j\not=j^{\prime}$
and $j^{\prime}\not=k^{\prime}$? It seems that we ran out o
possibilities of restricting the representations. Generically this
equations will not be satisfied because an intertwiner that
has simple $\iota$ in one tree decomposition has not only simple
$\iota^{\prime}$'s components in a different tree decomposition
and the equation would be violated. However there is a case in which
this happens trivially, namely when the dimension of the invariant
part of the tensor product of the four corresponding
representations is unity. Let us
write this condition as an equation since this corresponds to the
solution of the remaining 5 independent conditions, namely
\begin{equation}\label{din}
{\rm dim}\left({\rm
Inv}\left[\rho_{ij}\otimes\rho_{ik}\otimes\rho_{im}\otimes\rho_{ip}\right]\right)=1
\end{equation}
In this case $\iota$ would be simple in any tree decomposition
if it is simple in one particular one. Notice that this is the only solution
to our constraints as a trivial consequence of the theorem proven by Reisenberger
in \cite{reis3}. Our set of solutions are contained in the Barret-Crane solutions since our
intertwiners agree with the BC one every time that equation
(\ref{din}) is satisfied. Solutions to the previous equation can
be characterized as follows. Since all the $\rho$'s are simple we
can concentrate on their right (or left) components. Assume
$j_1\le j_2 \le j_3 \le j_4$ then the condition is
$j_1+j_2+j_3=j_4$. Explicitly, a few examples of solutions are
$(\frac{1}{2},\frac{1}{2},2,3)$,
$(\frac{3}{2},\frac{1}{2},\frac{5}{2},\frac{1}{2})$,
$(1,5,226,220)$, etc. The intertwiner color is completely
determined by the face colors and a choice of tree decomposition.
If we pair $j_1$ and $j_2$ in our previous example then
$\iota=j_1+j_2=j_4-j_3$. The amplitude $4SIM_{const}$ is independent of
the tree decomposition chosen.

\subsection{Gluing 4-simplexes}

Once we have solved equations (\ref{XX}) for a single 4-simplex we
can calculate the amplitude of any simplicial decomposition of
${\cal M}$, $\Delta$. This is achieved by putting together 4-simplexes with
consistent boundary connections and gluing them together by means
of integrating over the boundary data in the standard way.

Let us point out that there is a potential ambiguity in this step.
We have implemented constraints in the path integral and this
generally should be supplemented with the appropriate modification
of the measure. This could affect the values of lower dimensional
simplexes such as face and edge amplitudes. Constraints (\ref{XX})
act on each edge (tetrahedron) separately, heuristically one would
expect a Jacobian factor to modify the edge amplitude of the model
since the constraints are non linear functions of the $B$'s. A
rigorous derivation of such factors from the path integral
definition is lacking in our argument. We believe that this might
shed light on the problem of the correct normalization of this type
of spin foam models. A more detailed study of this issue is being
explored in \cite{maa}.

If we do so then we end up with
\begin{eqnarray}\label{gr4}
Z_{const}(\Delta)=\sum \limits_{ {\cal C}_f:\{f\} \rightarrow \rho^s_f }  \ \prod_{f \in \Delta^*} \Delta_{\rho} \prod_{e \in \Delta^*} A_e \prod_{v \in
{\Delta^*}}
\begin{array}{c}
\includegraphics[width=3cm]{4simm.eps}\end{array},
\end{eqnarray}
where $\rho^s_f$  denote the set of representations
selected by conditions (\ref{XX}), $\iota^{s}_{e}$ are the corresponding
colors of intertwiners, and $A_e$ is the appropriate edge amplitude
(undetermined in our prescription).

\subsection{Volume}

In this section we discuss a rather puzzling feature of the
model we have defined above.

If we consider boundaries, then the spin-network states induced as
boundaries of spin foams are four-valent and the representations
of the corresponding edges satisfy (\ref{din}). Now using the
standard definition of the volume operator on this set of states
we obtain an identically zero result, i.e, the 3-volume operator
$V_{(3)}$ annihilates the states that solve (\ref{XX}). The reason
is that the volume is given by\cite{barb2}
\begin{equation}
V^{2}_{(3)} \propto \left[({\cal X}_i+{\cal X}_j)^2,({\cal X}_i+{\cal X}_k)^2\right],
\end{equation}
where the square is taken using the internal metric $\delta_{IJ}$.
The solutions to the constraints happen to diagonalize both
operators in the commutator which implies $V_{(3)}=0$.

\section{Degenerate sector}\label{ddeegg}

As shown in \cite{reis6, fre6}, constraints (\ref{constraints})
correspond to the non-degenerate phase of solutions of the general
constraints (i.e., phase with $e\not=0$). In \cite{reis6}
Reisenberger explicitly solved the constraints in the degenerate
sectors and showed that, in these cases, the action reduces to
\begin{equation}\label{rei}
S^{\pm}_{deg}=\int B^{r}_i\wedge (F_i(A^r) \pm V_i^j F_j(A^{\ell})),
\end{equation}
where the upper index $r$ (respectively $\ell$) denotes the
self-dual (respectively anti-self-dual) part of $B$ and $A$ in the
internal space, and $V \in SO(3)$.

Let us concentrate in the sector with the minus sign in the
previous expression. Then it is straightforward to define the
discretized path integral along the same lines as BF theory in
Section \ref{BF}. The result is
\begin{equation}\label{Zdeg}
{\cal Z}(\Delta)=\int \prod_{f \in \Delta^*} dB^{r\va (3)}_f dv_f \
\prod_{e \in \Delta^*} dg^{\ell}_e dg^{r}_e
\ e^{i {\rm Tr}\left[B^{r}_f U^{r}_f v_f U^{\ell - 1}_f v_f^{-1}\right]}.
\end{equation}
Integrating over the $B$ field we obtain
\begin{equation}\label{deg}Z(\Delta)= \int \prod_{e \in \Delta^*}
dg^{\ell}_e dg^{r}_e  \prod_{f \in \Delta^*} dv_f\
\delta^{(3)}(g^{r}_{e_1} \cdots g^{r}_{e_n}v_f(g^{\ell}_{e_1}
\cdots g^{\ell}_{e_n})^{-1}v_f^{-1}),
\end{equation}
where $dg^{\ell}_e$, $dg^{r}_e$, and $dv_f$ are defined in terms
of the $SU(2)$ Haar measure and the delta function $\delta^{(3)}$
denotes an $SU(2)$ distribution.

\begin{figure}[h]
\centering{\includegraphics[width=5cm]{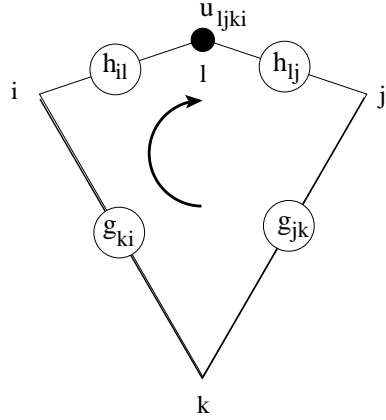}}
\caption{Diagramatic representation of the single wedge contribution (\ref{wedgea}).
The group variables $g_{ki}$ and $g_{jk}$ correspond to the internal connections while
$h_{il}$ and $h_{lj}$ to boundary data. $u_{ljki}$ is an independent auxiliary variable in the $SU(2)$
subgroup.}\label{wedge}
\end{figure}

\begin{figure}[h]
\centering{\includegraphics[width=5cm]{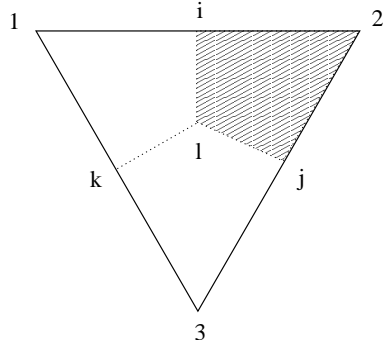}}
\caption{A triangular face made up of three wedges. The wedge $2ilj$
has been emphasized.}\label{facewedge}
\end{figure}

In order to obtain the corresponding state-sum it is easier to
concentrate on a single 4-simplex amplitude.
Furthermore, we start by the wedge shown in Figure
\ref{wedge}. In this figure we represent one of the 10 wedges that form
a 4-simplex atom (see Figure \ref{chunk}). Both the internal connection $g_{ij}$ ($g_{ij}=g^{-1}_{ji}$)
and the boundary connection variables $h_{ij}$ ($h_{ij}=h^{-1}_{ji}$) are in $Spin(4)$, while
$u_{ljki} \in SU(2)\subset Spin(4)$ is an auxiliary variable. The $SU(2)$ subgroup is
defined as the diagonal insertion $ug=(ug^{\ell}, ug^r)$.
The wedge amplitude is defined as
\begin{equation}\label{wedgea}
w=\int du_{ljki} \ \delta^{(6)}(g_{ki}h_{il}u_{ljki}h_{lj}g_{jk})
\end{equation}
according to the notation in Figure \ref{wedge} and where the
$\delta^{(6)}$ denotes a $Spin(4)$ delta distribution.
Any face in the 2-complex will be defined by as many such wedges
as 4-simplexes share the corresponding face. Figure
\ref{facewedge} illustrates the case for a triangular face. The
vertices $1,2$ and $3$ correspond to the centers of the three
4-simplexes sharing the face. The dotted line denotes the region
along which the boundary of the three atoms (Figure \ref{chunk})
join.

It is easy to check that integrating over all but one
boundary variables $h_{ij}$, the contribution of a
combination of wedges forming a face $f\in \Delta^*$
is given by
\begin{equation}
\int du_f dh\ \delta^{(6)}(U_fh u_f h^{-1}),
\end{equation}
where $U_f\in Spin(4)$ is the discrete holonomy, $u_f\in SU(2)\subset Spin(4)$ is a product of the $u_w$
associated to the corresponding wedges and $h$ is the remaining boundary connection.
In the case shown in
Figure \ref{facewedge}, $U_f=g_{1i}g_{i2}g_{2j}g_{j3}g_{3k}g_{k1}$, $u_f=u_3u_2u_1$, and $h=h_{kl}$.
Using that $Spin(4)=SU(2)\times SU(2)$ and the definition of the $SU(2)$
subgroup where $u$ lives, the integral over $u_f$ of the previous equation becomes
\begin{eqnarray}&& \nonumber
\int du_f dh^{\ell} dh^{r} \ \delta^{(3)}(U^{\ell}_fh^{\ell} u_f
h^{\ell -1}) \delta^{(3)}(U^{r}_fh^{r} u_f h^{r -1})
\\ &&=\int d(h^{\ell}h^{r-1}) \  \delta^{(3)}(U^{\ell}_fh^{\ell}h^{r -1}U^{r -1}_f h^{r}h^{\ell -1}),
\end{eqnarray}
where we have used that $dh=dh^{\ell}dh^r$ and
$\delta^{(6)}(g)=\delta^{(3)}(g^{\ell})\delta^{(3)}(g^{r})$ as
well as the invariance and normalization of the Haar measure. The
previous face amplitude coincides with that in (\ref{deg}) if we
define $v_f=h^{\ell}h^{r -1}$. Therefore, (\ref{wedgea}) defines
the wedge amplitude of (\ref{deg}).

Now we can write the analog of equation (\ref{bfwf}) for the
4-simplex amplitude, $4SIM_{Deg}(h_{ij})$, putting together the 10 corresponding wedges and
integrating over the internal $g$'s, namely
\begin{equation}\label{degwf}
4SIM_{Deg}(h_{ij})=\sum \limits_{\rho_1 \cdots \rho_{10}}\ \ \sum
\limits_{\iota_1 \cdots \iota_5}
\begin{array}{c}
\includegraphics[width=2cm]{4simm.eps}\end{array}
\begin{array}{c}
\includegraphics[width=4.5cm]{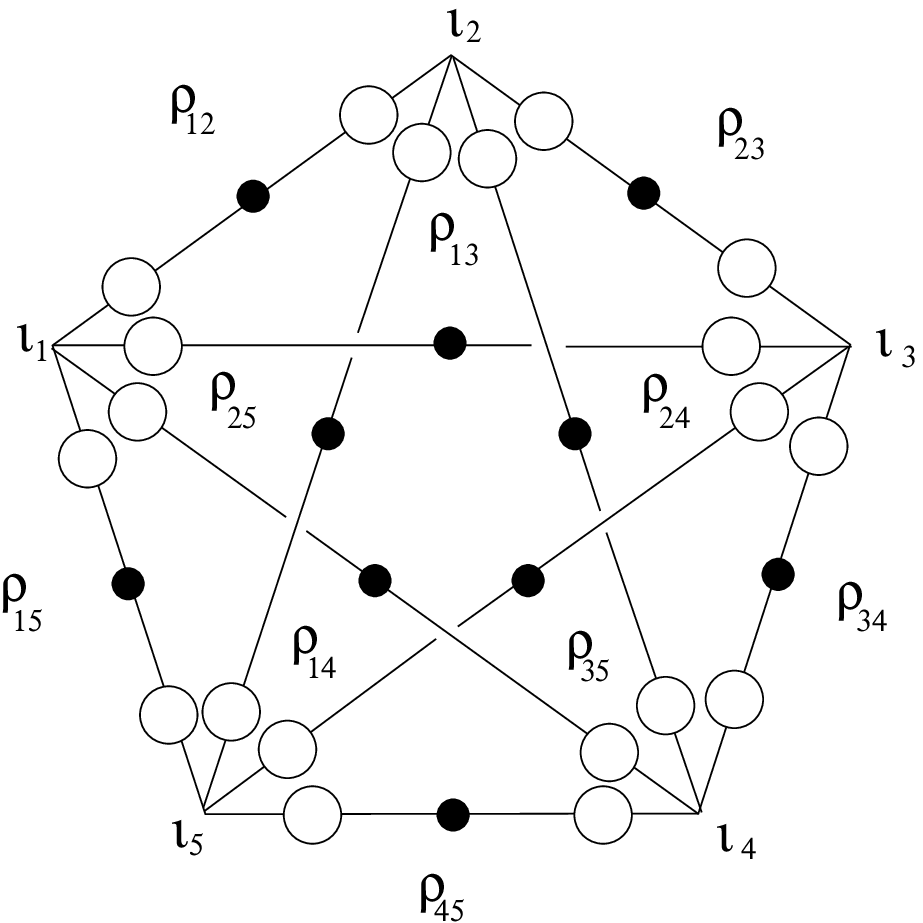}\end{array},
\end{equation}
where the dark dots denote integration over the $SU(2)$ diagonal
subgroup, and the white circles represent the boundary
connections. To keep the diagrammatic notation simple we have
dropped some labels. The next step is to perform the integration
over the $u$'s. We concentrate on a single intertwiner in
(\ref{degwf}), i.e., a single node in the pentagonal diagram on
the right of the previous equation. Using the orthogonality of
$SU(2)$ unitary irreducible representations and the fact that the
representations $\rho$ of $Spin(4)$ are of the form $\rho=j\otimes
k$ for $j,k$ $SU(2)$ unitary irreducible representations we have
\begin{equation}\label{uuu}\!\!\!\!\!
\begin{array}{c}
\includegraphics[width=4.7cm]{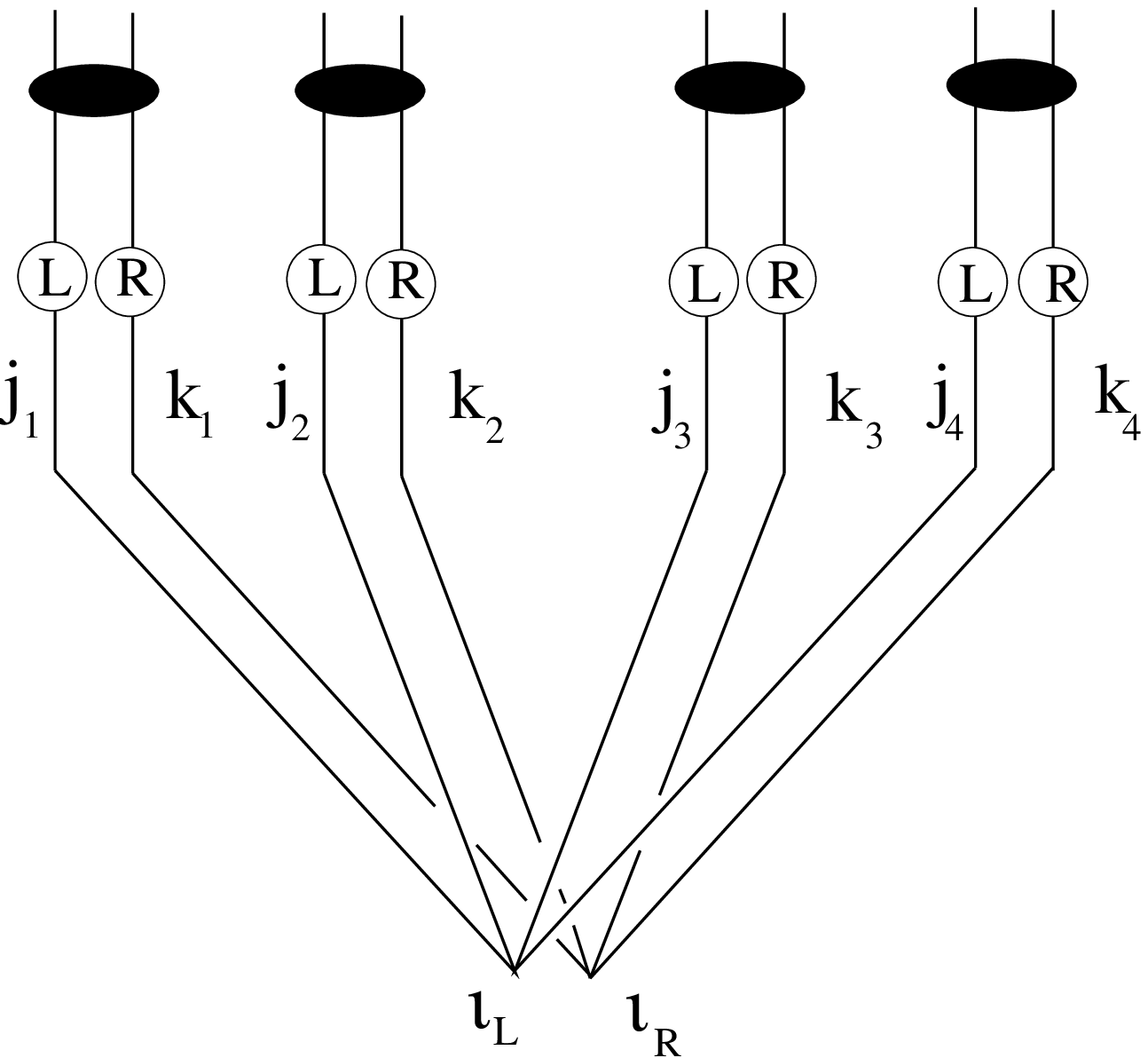}\end{array}\!\!\!\!\!\!\!\!\!\!=
\frac{\delta_{j_1,k_1}\cdots \delta_{j_4,k_4}}{(2j_1+1)\cdots(2j_4+1)}\!\!\!\!\!\!\!\! \begin{array}{c}
\includegraphics[width=4.2cm]{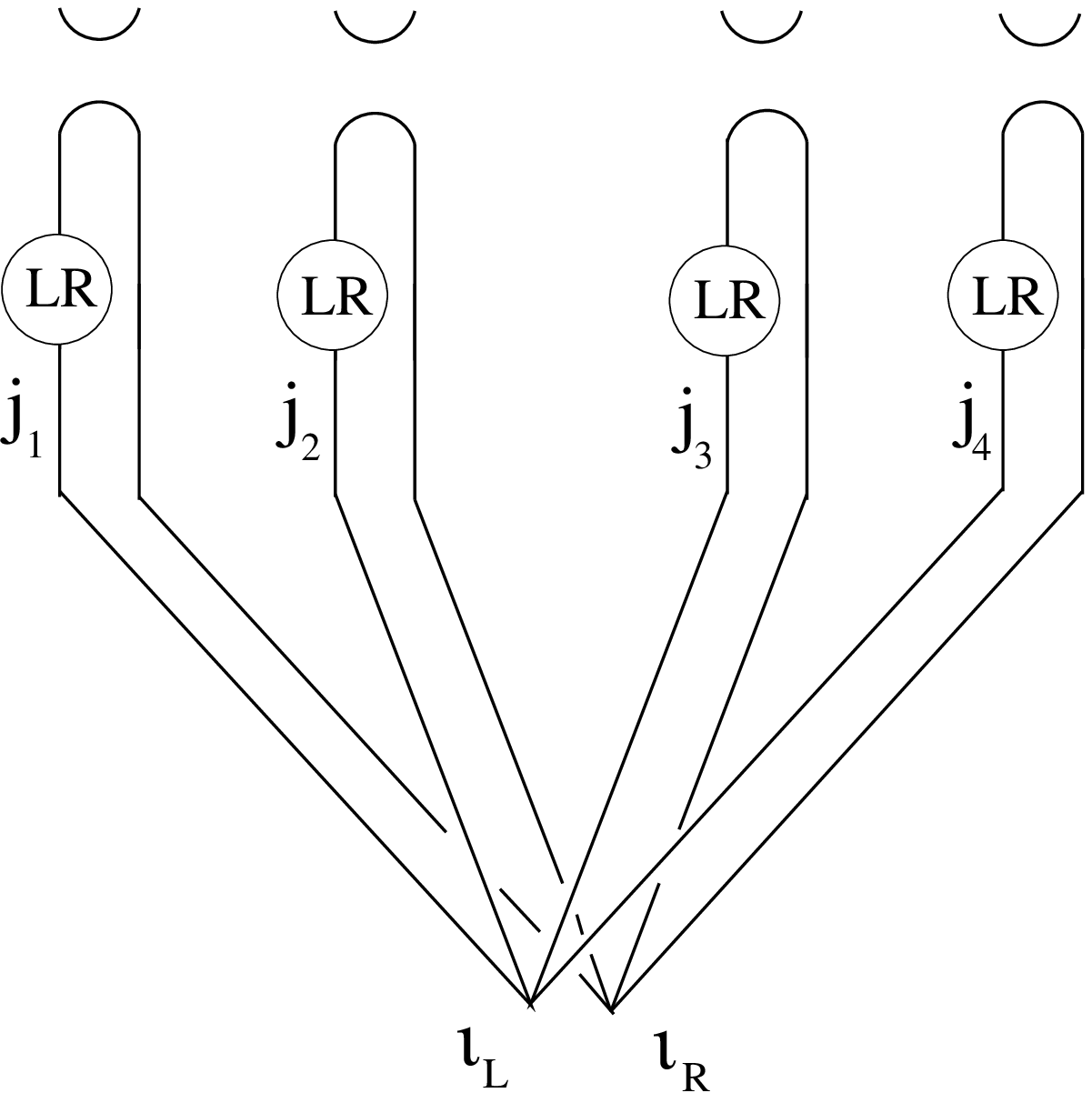}\end{array},
\end{equation}
where we represent representations $\rho=j\otimes k$ as parallel lines,
the symbol $R$, $L$ in the circles on the left denotes $h^r$, $h^{\ell}$
$SU(2)$-representation matrices, $RL$ on the right denotes the
product $h^{r}h^{-1\ell}$, and the dark dots are subgroup integrations.

The Kronecker deltas in the previous equation implies that the
$\rho$'s labeling faces must be simple, i.e., $\rho=j\otimes j$.

Finally, it is easy to verify that when gluing various 4-simplex
atoms together by means of integrating over matching boundary
connections the integration simply set $\iota_R=\iota^{\prime}_R$,
$\iota_L=\iota^{\prime}_L$ where $\iota$, $\iota^{\prime}$ are the
intertwiners corresponding to the tetrahedron shared by the two
4-simplexes. Notice that no simplicity condition is imposed on
$\iota$. Now consider an arbitrary face bounded by $n$ edges. Such
a face is made up $n$ wedges. Therefore there is a factor
$\Delta_{jj}^{2n}=({2j+1})^{2n}$ coming from the delta function
mode expansion in (recall Footnote (\ref{PW}).), a factor
$({2j+1})^{-n}$ from the factors in (\ref{uuu}), and finally a
factor $({2j+1})^{-n}$ from the boundary connection integrations
in the gluing. This results in a face amplitude equal to unity.

Putting all this together one gets a spin foam model were only face representations
are constrained to be simple while intertwiners are arbitrary.
Explicitly
\begin{eqnarray}\label{grdeg}
Z_{deg}(\Delta)=\sum \limits_{ {\cal C}_f:\{f\} \rightarrow
\rho^{s}_f }  \sum \limits_{{\cal C}_e:\{e\} \rightarrow \{
\iota_e \}} \ \prod_{v \in {\Delta^*}}
\begin{array}{c}
\includegraphics[width=3cm]{4simm.eps}\end{array}.
\end{eqnarray}

This is precisely the spin foam obtained in \cite{fre2}! This
model was obtained as a natural modification of the GFT that
defines a variant of the BC model. Here we have rediscovered the
model from the systematic quantization of $S^{-}_{deg}$ defined in
(\ref{rei}). This establishes the relation of the model with a
classical action! It corresponds to spin foam quantization of the
`$-$' degenerate sector of $SO(4)$ Plebanski's theory.

The $+$ sector action (\ref{rei}) can be treated in a similar way.
The only modification is that of the subgroup. Instead of using
the diagonal insertion defined above one has to define $u\in
SU(2)\subset Spin(4)$ so that $ug=(ug^{\ell}, u^{-1}g^r)$.

We have restricted to simplicial decompositions but all this
should be generalizable along the lines of reference \cite{a1} for
arbitrary cellular decompositions of $\cal M$. This generalization
seems straightforward although it should be investigated in
detail.

To conclude this section let us notice that the allowed 4-simplex
configurations of the model of Section \ref{ccc} are fully
contained in the set of 4-simplex configurations of the model
obtained here. We come back to this issue in the following
section.

\section{Discussion}\label{Diss}

The principal idea behind this work was to study the spin foam
quantization of Plebanski formulation of gravity by restricting
the paths that appear in the $SO(4)$ BF theory. This strategy is
supported by the fact that Plebanski's action can be thought of as
the $SO(4)$ BF theory, supplemented by certain constraints on the
B field. Gravity in the Palatini formulation is obtained as one of
the non-degenerate sectors of the solutions to the classical
constraints. In the model introduced in section \ref{ccc} we
defined a prescription for implementing these constraints by
restricting the set of histories of the BF theory to those
satisfying the `quantum analog' of (\ref{constraints}). Solution
configurations of a single 4-simplex in the model are special
4-simplex BF configurations. Even though the 4-simplex
configurations appearing here are a sub-set of the Barrett-Crane
configurations, the great majority of the Barrett-Crane
configurations (independently of the normalization chosen) are
excluded by the requirement that they be BF configurations. The
nature of the constraints in the BC model is essentially the same
as the ones defined here. The difference is that in the BC case
constraints are implemented on a single intertwiner while here we
keep track of the fact that intertwiners appear in pairs in the BF
state sum (see (\ref{4dp})).

There are alternative ways to motivate the definition of the Barrett-Crane
model which are independent of the line of thought used here.
There is also evidence that relate it to a theory of quantum
gravity. However, we believe that this work shows
that there is no obvious way to interpret it as the
quantization of Plebanski's action. Using reasonable definitions
we have shown that one obtains a more
restrictive state sum.

In the context of the BC model,
reference \cite{baez6} shows how one can restrict the states of the
`quantum tetrahedron' so that fake tetrahedra are ruled out
of the state sum. In our context this amounts to resolving the ambiguity
between the $e\wedge e$ and ${}^*(e\wedge e)$ solutions of the constraints (see (\ref{ambi})) at the
quantum level. In \cite{baez6} it is shown that the `correct' configurations are selected
by imposing the so-called chirality constraint which is automatically satisfied at the quantum
level because it can be written as the commutator of the simplicity
constraints (\ref{XX}). It is also shown that the two spaces of solutions
($e\wedge e$ and ${}^*(e\wedge e)$ respectively) intersect on the set of
configurations for which $V_{(3)}=0$.
It is easy to see that all this can be translated to our context.
The vanishing of the volume operator implies that in our model one
can not distinguish the two type of configurations and that the ambiguity (\ref{ambi})
remains at the quantum level.

The model of Section \ref{ccc} contains only degenerate
configurations in the sense that spin-network states on the
boundaries have zero volume. This shows that the model cannot
reproduce any of the semi-classical states of general relativity.
Somehow our definition of the constraints at the level of the
state-sum are so strong that non-degenerate configurations have
been eliminated. Some evidence supporting this view can be
obtained considering the following argument. Constraints are
implemented locally on each 4-simplex; therefore, we can
concentrate on a single 4-simplex to analyze their action. If we
do so, then we conclude that all the 4-simplex configurations in
the model (\ref{gr4}) are entirely contained in the set of
4-simplex configurations of the model found in Section
\ref{ddeegg}.

Is there a way out? If we maintain the point of view of defining
the model starting from Plebanski's action then the problem can be
traced back to our definition of constraints. As it was pointed
out in \cite{fre6} there are two ways to write Plebanski's
constraints in the non-degenerate sector (i.e., when $e\not=0$ in
(\ref{constraints})).
If we stick with (\ref{constraints}) then one can try to change
the definition of the $B$ operators. One possibility would be to
change right-invariant by left invariant vector fields in the
definition (\ref{B_f}). One can do this consistently only if one
changes the orientation of the 4-simplex in which case the final
result remains the same. One can try to use the sum of
right-invariant and left invariant vector fields. This is
certainly a possibility, and actually converges faster to the
value of $B$ in the sense of equation (\ref{RIV}) when
$U\rightarrow 1$. However, since the constraints are quadratic in
the $B$ there will be cross-terms in (\ref{XX}). This terms cannot
be expressed in terms of $Spin(4)$ Casimir operators, and
consequently, the constraints cannot be solved in terms of simple
restrictions on the set of representations involved in the
state-sum. There seem to be no obvious way to use
(\ref{constraints}) and avoid the discouraging results of Section
\ref{ccc}.

The other possibility is to use
\begin{equation}\label{new}
\epsilon^{\mu\nu\rho\sigma}B^{IJ}_{\mu\nu}B^{KL}_{\rho\sigma}\propto \epsilon^{IJKL},
\end{equation}
where the difference with (\ref{constraints}) is in the fact that
we have traded internal with space-time indices. However with this
choice, the connection with the BC model becomes much more
uncertain. Notice that in this form, the constraints have free
Lie-algebra indices and therefore cannot be written as Casimir
operators as they stand. This version of the constraints has been
studied in the literature. Such constraints have been incorporated
in a spin foam model of Riemannian general relativity in terms of
self-dual variables by Reisenberger in \cite{reis6}. In the
context of $SO(4)$ Plebanski's action a model along this lines has
been defined by Freidel et al. in \cite{fre5}. But all these
models are quite different from the BC model. We believe that this
shows that there is no obvious means of interpreting the BC model
as a spin foam quantization of Plebanski's theory.

Let us conclude by analyzing the results of the last section. In
Section \ref{ddeegg} we quantized the degenerate sectors of
Plebanski's action in a fairly straightforward way. In this case
we do not impose any constraints and the state sum follows
directly from the discretized definition of the path integral of
the theory. There are no ambiguities in lower dimensional
simplex-amplitudes. The model turns out to be precisely the one
introduced by De Pietri, Freidel, Krasnov and Rovelli in
\cite{fre2}. This work establishes a clear connection between that
model and the effective action corresponding to one of the
degenerate sectors of Plebanski's action.

Finally the model is well defined, is not topological and has a
clear connection to a continuous action. It is somehow between the
theory we want to define and the simpler theories we understand
well but do not have local excitations (such as BF theory and
gravity in lower dimensions). From this viewpoint we believe that
it might  be useful to explore its properties as a `toy model' for
understanding open issues in the spin foam approach to quantum
gravity. Among these is the very important problem of the
continuum limit (i.e., the issue of summing-over versus refining discretizations) and
the interpretation of the path integral in the diffeomorphism
invariant context (time evolution versus the projector/extractor operator on  physical states).

\section{Acknowledgments}
I thank  Abhay Ashtekar, Martin Bojowald and Amit Ghosh for
stimulating discussion and insightful questions. I also thank John
Baez, Rodolfo Gambini, Jorge Pullin, Michael Reisenberger and
Carlo Rovelli for motivating remarks. I am grateful to Abhay for
many suggestions that have improve the presentation of this
manuscript. This work was supported in part by NSF Grant
PHY-0090091, and Eberly Research Funds of Penn State.

\end{document}